\documentclass[aps,pre,twocolumn,groupedaddress]{revtex4-2}
\usepackage{graphicx}
\usepackage{amsmath,amssymb}
\usepackage[normalem]{ulem}
\usepackage{xcolor}

\begin{document}

\title{Helicity modulus and chiral symmetry breaking for boundary conditions with finite twist}

\author{Gaurav Khairnar, Thomas Vojta}
\email[]{grktmk@mst.edu, vojtat@mst.edu}
\affiliation{Department of Physics, Missouri University of Science and Technology, Rolla, MO, USA, 65401}

\begin{abstract}
We study the response of a two-dimensional classical XY model to a finite (non-infinitesimal) twist of the boundary conditions.
We use Monte Carlo simulations to evaluate the free energy difference between periodic and twisted-periodic boundary conditions
and find deviations from the expected quadratic dependence on the twist angle.
Consequently, the helicity modulus (spin-stiffness) shows a non-trivial dependence on the twist angle.
We show that the deviation from the expected behavior arises because of the mixing of states with opposite chirality
which leads to an additional entropy contribution in the quasi-long-range ordered phase.
We give an improved prescription for the numerical evaluation of the helicity modulus for a finite twist, and we discuss the spontaneous breaking of the chiral symmetry for the
anti-periodic boundary conditions.
We also discuss applications to discrete spin systems and some experimental scenarios where boundary conditions with finite twist are necessary.
\end{abstract}
\date{\today}
\maketitle

\section{Introduction}
\label{sec:Introduction}
The study of magnetic materials with the help of spin models such as the Ising or Heisenberg models has resulted in tremendous progress in condensed matter physics.
For clean systems with short-range interactions, surface energies are negligible, and the choice of boundary conditions does not affect the thermodynamic bulk behavior.
On the other hand, in disordered systems such as spin glasses, effects of various boundary conditions have been a topic of active interest \cite{banavar1982influence,newman1992multiple,endo2021roles}.
In the presence of long-range interactions in an Ising spin system, different choices of boundary condition lead to seemingly different thermodynamic behaviors. By choosing artificial coupling-dependent boundary condition, an uncountable number of exotic spin states of the ground state can be generated at any temperature, whereas free boundary conditions, which are considered physical, do not generate the same effect \cite{gandolfi1993exotic}. In XY or Heisenberg spin glasses, small rotations at the boundary are expected to yield non-smooth changes in the ground state \cite{fisher1988equilibrium}.
In these systems, different boundary conditions are advantageous for revealing the numerous phases and their physical properties \cite{akino2002,kawamura1987chiral,ney1995chiral}.

Even in system in which the boundary conditions do not affect the bulk behavior, they can have interesting and observable consequences. For example, thermodynamic Casimir forces \cite{FisherdeGennes78}, which arise due to the confinement of critical order parameter fluctuations, were found to be affected by the boundary conditions \cite{BloteCardyNightingale86,Cardy86b,Affleck86,bergknoff2011,dantchev2017} (for a recent review of exact results see Ref.\ \cite{dantchev2023critical}). An analogous force, the Helmholtz force, arises in the canonical ensemble and also shows a strong dependence on the boundary conditions \cite{dantchev2024casimir}.
Moreover, the response of a system to a change in boundary conditions can be employed as a tool to measure (bulk) equilibrium properties.

In this paper, we reconsider the seemingly simple but surprisingly complex effects of a finite (non-infinitesimal) twist of the boundary conditions on a ferromagnetic XY system. Brown and Ciftan \citep{BrownHelicity} suggested, in the context of the three-dimensional Heisenberg model, that the mixing of states with differing chirality plays an important role  and affects the observed helicity modulus.
However, a quantitative understanding of this effect and its origins has not been achieved yet.
We therefore study this question in detail for the two-dimensional classical XY model by means of large-scale Monte Carlo simulations.

We find that the free energy cost of a non-infinitesimal twist in the boundary condition in the quasi long-range ordered (QLRO) low-temperature phase deviates from the expected quadratic dependence on the twist angle.
Mixing of states with opposite chirality provides an extra entropy contribution which takes the value $\ln(2)$ (in units in which $k_B=1$) for a $\pi$-twist.
Beyond their intrinsic interest, our results potentially apply to experiments aimed at detecting the Berezinskii--Kosterlitz--Thouless (BKT) transition \cite{Berezinskii71,KosterlitzThouless73}.
They are also important for systems with discrete $Z(N)$ (clock) order parameter symmetry where any twist in the boundary conditions is necessarily non-infinitesimal. In addition, our findings enable novel numerical algorithms for computing the helicity modulus in simulations with a finite twist.

Our paper is organized as follows. In section \ref{sec:model}, we introduce the XY Hamiltonian under boundary conditions with an arbitrary twist, and we define the helicity modulus.
Section \ref{sec:MC} contains the details of the numerical simulations. In section \ref{sec:results}, we present our results for the dependence of the free energy on the twist angle and discuss the resulting
helicity modulus. We also analyze the spontaneous breaking of the chiral symmetry that occurs for antiperiodic boundary conditions below the BKT transition.
We conclude in section \ref{sec:conclusions}.

\section{The Model}
\label{sec:model}

We are interested in the classical XY model, a system of planar spins described by the Hamiltonian
\begin{equation}
\label{eq_hmltn}
H = -J \sum_{<ij>} \mathbf{S}_i \cdot \mathbf{S}_j = -J \sum_{<ij>} \cos(\phi_i - \phi_j)~.
\end{equation}
Here, $J>0$ denotes the ferromagnetic exchange interaction (which will be set to unity in the simulations),
the sum is over pairs of nearest neighbors on a
$d$-dimensional hypercubic lattice, and $\mathbf{S} =(S_x,S_y)$ is a two-component unit vector.
Equivalently, the XY spins can be represented by their phases $\phi$, defined via $S_x = \cos \phi, ~ S_y= \sin \phi$.
Twisted boundary conditions can be implemented by fixing the spins at two opposite boundaries at specific orientations (phases)
with a fixed angle $\Theta$ between them.
Alternatively, we consider the Hamiltonian (\ref{eq_hmltn}) with periodic boundary conditions and
modify the interactions across one of the boundaries to introduce the twist.
This is achieved by replacing the interaction terms across the chosen boundary by $-J \cos (\phi_i -\phi_j -\Theta)$.
We call these boundary conditions twisted-periodic,  and the usual periodic boundary conditions are recovered for $\Theta=0$.

The phase diagram of the classical XY model is well-understood. Long-range order is impossible in one and two dimensions at any
nonzero temperature due to the Mermin-Wagner theorem \cite{MerminWagner66}. In three and higher dimensions, there is a phase
transition between a paramagnetic high-temperature phase and a ferromagnetic low-temperature phase. The two-dimensional XY model
is special because the system undergoes a BKT phase transition into a quasi long-range ordered low-temperature phase.

In a long-range ordered or quasi long-range ordered phase, a twist in the boundary conditions increases the system's
free energy. The system can lower its free energy by distributing the total twist (angular difference) $\Theta$ over the
entire sample, i.e., by gradually changing the average orientation of the spins in the bulk. For a system of linear size $L$,
the lowest free energy is expected when the average phase changes by $\Theta/L$ between neighboring sites (in the direction the
twist is applied). For large $L$, this local phase change is small, which suggests that the free energy can be expanded in
powers of $\Theta/L$. As the free energy difference $\Delta F$ between the twisted and untwisted systems must be an even
function of $\Theta$, the expansion is expected to take the form
\begin{equation}
\label{eq_freediff}
\Delta F = F_\Theta - F_0 = \frac{\rho_s}{2} \left( \frac{\Theta}{L} \right) ^2 L^d
\end{equation}
to quadratic order in $\Theta/L$. This relation can be understood as a definition of the helicity modulus (or spin stiffness) $\rho_s$,
\begin{equation}
\label{eq_stiff}
\rho_s(\Theta) = \frac{2 \Delta F}{\Theta^2} L^{2-d}~.
\end{equation}
This definition still depends on the value of the imposed twist angle $\Theta$. Some papers in the literature including the
seminal work by Fisher, Barber, and Jasnow \cite{FisherBarberJasnow73} define the helicity modulus via a twist angle of $\pi$, i.e.,
via the free energy difference between periodic and antiperiodic boundary conditions (in analogy with the study of interfacial energies in Ising models).
Other authors define the helicity modulus via the response to an infinitesimal twist (in the spirit of linear response theory),
\begin{equation}
\label{eq_stiff_inf}
\rho_{s0} = \rho_s(0) = \left(\frac{\partial^2 F}{\partial \Theta^2}\right)_{\Theta=0} L^{2-d}~.
\end{equation}
It has generally been assumed that the two definitions lead to the same stiffness values because the local, layer-to-layer twist $\Theta/L$ is small in either case (in the thermodynamic limit),
justifying the expansion (\ref{eq_freediff}) of the free energy  \cite{RudnickJasnow77,BanavarJasnow78}. However, we will see that this is not the case, at least not in two space
dimensions.

The definition (\ref{eq_stiff_inf}) has the advantage that the second derivative of the free energy can be expressed in terms of appropriate correlation functions of
the untwisted system. This leads to the formula \cite{Teital1983,Caffarel_1994}
\begin{eqnarray}
&\left(\frac{\partial^2F}{\partial\Theta^2} \right)_{\Theta=0} = \frac{1}{L^{2}} \sum_{<ij>} J \left\langle \cos(\phi_i-\phi_j) \right\rangle(x_i-x_j)^2  \nonumber \\
                                  & -\frac{\beta}{L^{2}} \left\langle \left\lbrace \sum_{<ij>} J\sin(\phi_i-\phi_j)(x_i-x_j) \right\rbrace^2 \right\rangle
\label{eq:infinitesimal_stiffness}
\end{eqnarray}
where $x_i$ is the coordinate of site $i$ in the twisted direction, and $\langle ... \rangle$ denotes the thermodynamic average evaluated at $\Theta=0$. For a derivation of Eq.\ (\ref{eq:infinitesimal_stiffness}), see Appendix \ref{sec_appndx}.
This formula allows the evaluation of the helicity modulus without actually having to apply twisted boundary conditions.

As the paramagnetic phase is insensitive to the boundary conditions, the free energy difference decays exponentially with system size, $\Delta F \sim e^{-L/\xi}$ , where $\xi$ is the correlation length. This implies $\rho_s=0$ in the paramagnetic phase in the thermodynamic limit. In contrast, $\Delta F$ is expected to scale as $L^{d-2}$ in an ordered or quasi long-range ordered phase, and the helicity modulus is finite. In two dimensions, $\rho_s$ is known to have an universal jump at the BKT phase transition. For $d>2$, $\rho_s$ vanishes continuously at $T_c$, governed by the critical behavior of the phase transition. In the rest of the paper, we focus on two dimensions, but we will comment on higher dimensions in the concluding section.

\section{Numerical Simulations}
\label{sec:MC}
We perform large-scale Monte Carlo simulations to evaluate the free energy difference between systems with
periodic and twisted-periodic boundary conditions. The free energy cannot be measured directly in a standard
Monte Carlo simulation. Instead, it can be evaluated explicitly by integrating the internal energy $U = \langle H \rangle$
over the inverse temperature $\beta=1/T$,
\begin{equation}
\label{eq_freeint}
F(T) = F(T_0) + T \int_{\beta_0}^{\beta} d\beta ' U(\beta ')~.
\end{equation}
We choose the starting temperature $T_0$ sufficiently high (well above the BKT transition) such that the free energies of the twisted and untwisted systems agree with each other within the statistical errors. This ensures that $F(T_0)$ drops out of the free energy difference (\ref{eq_freediff}). We note that there are alternative approaches that allow one to directly measure the free energy difference between different boundary conditions in a simulation. This is achieved by including appropriate boundary terms as dynamical variables in the Monte Carlo scheme (see, e.g., Refs.\ \cite{Hasenbusch93,Hukushima99}). The challenge in these approaches is to ensure a sufficiently rapid relaxation of the boundary variables, especially in the absence of cluster algorithms.

Our simulations employ both the single spin-flip Metropolis algorithm \cite{metropolis1949monte,MRRT53} and the Wolff cluster-flip algorithm \cite{wolff1989collective}. For systems with periodic boundary conditions, the efficient Wolff algorithm greatly reduces  critical slowing down. Thus, a full MC sweep consists of one Metropolis sweep followed by one Wolff sweep for the case of periodic boundary conditions. However, for twisted-periodic boundary conditions with twist angle $0 < \Theta < \pi$, the Wolff algorithm cannot be employed. This stems from the fact that the angle between two spins of a Wolff cluster is not preserved (but rather changes sign) when the cluster is flipped. Consequently, the cluster flip changes the energy of a twisted bond inside the cluster, invalidating the algorithm. In the case of twisted-periodic boundary conditions with $0 < \Theta < \pi$, we therefore only employ Metropolis sweeps.
For a twist angle of exactly $\pi$, the twisted bonds effectively become antiferromagnetic as $\cos(\Delta\phi - \pi)=-\cos(\Delta \phi)$. The energy of an antiferromagnetic bond is invariant under a sign change of $\Delta \phi$, and the Wolff algorithm can be used.

To facilitate the numerical integration (\ref{eq_freeint}) for the free energy, we initiate the simulations at the highest temperature $T_0$ (using a ``hot'' start, i.e., all spins are randomly oriented at the beginning of the simulation). The temperature is then reduced in small steps $dT$ until the desired final temperature is reached. Most production simulations started from $T_0=30$, much higher than the BKT transition temperature of $T_c=0.89290(5)$ \cite{ueda2021}. The temperature step $dT$ is gradually decreased from $dT=0.5$ at $T_0$ to $dT=0.02$ in the transition region and below. To check how sensitive the free energy difference $\Delta F$ is to these parameters, we performed tests with $T_0$ as high as 90 and $dT$ as low as 0.01. The free energy differences resulting from these test calculations agreed with the production results within their statistical errors.

For periodic boundary conditions, we perform up to $2000$ full equilibration sweeps (each consisting of a Metropolis sweep followed by a Wolff sweep) and up to $8000$ full measurement sweeps at each temperature step. For twisted-periodic boundary conditions, we perform up to about $80\,000$ (Metropolis only) equilibration sweeps and up to about $160\,000$ measurement sweeps. (As usual, the quality of the equilibration was confirmed by comparing the results of hot and cold starts.) We simulate systems of linear sizes up to $L^2 = 80^2$ and average the results over about $5000$ samples. Each sample is subjected to periodic and twisted-periodic boundary conditions, and the resulting free energies are compared to evaluate the helicity modulus from  eq.\ (\ref{eq_stiff}).

We use an ensemble method (see, e.g., Ref.\ \cite{khairnar2021phase}) to estimate the error of the free energy. We generate a large ensemble of synthetic internal energy curves ${U_i(T) = U(T) + r(T)~\Delta U(T) }$, where $\Delta U$ is the statistical error obtained from Monte Carlo and $r(T)$ is a random number chosen from a normal distribution of unit variance.
Integrating these curves via (\ref{eq_freeint}) generates an ensemble of free energies $F_i$.
Mean and standard deviation of this ensemble are then propagated through eq.\ (\ref{eq_stiff}) to find
the helicity modulus and the associated error.
\section{Results}
\label{sec:results}
\subsection{Free energy response to finite twist}
\label{sec:results-free-energy}
\begin{figure}[t]
\centering
\includegraphics[width=\columnwidth]{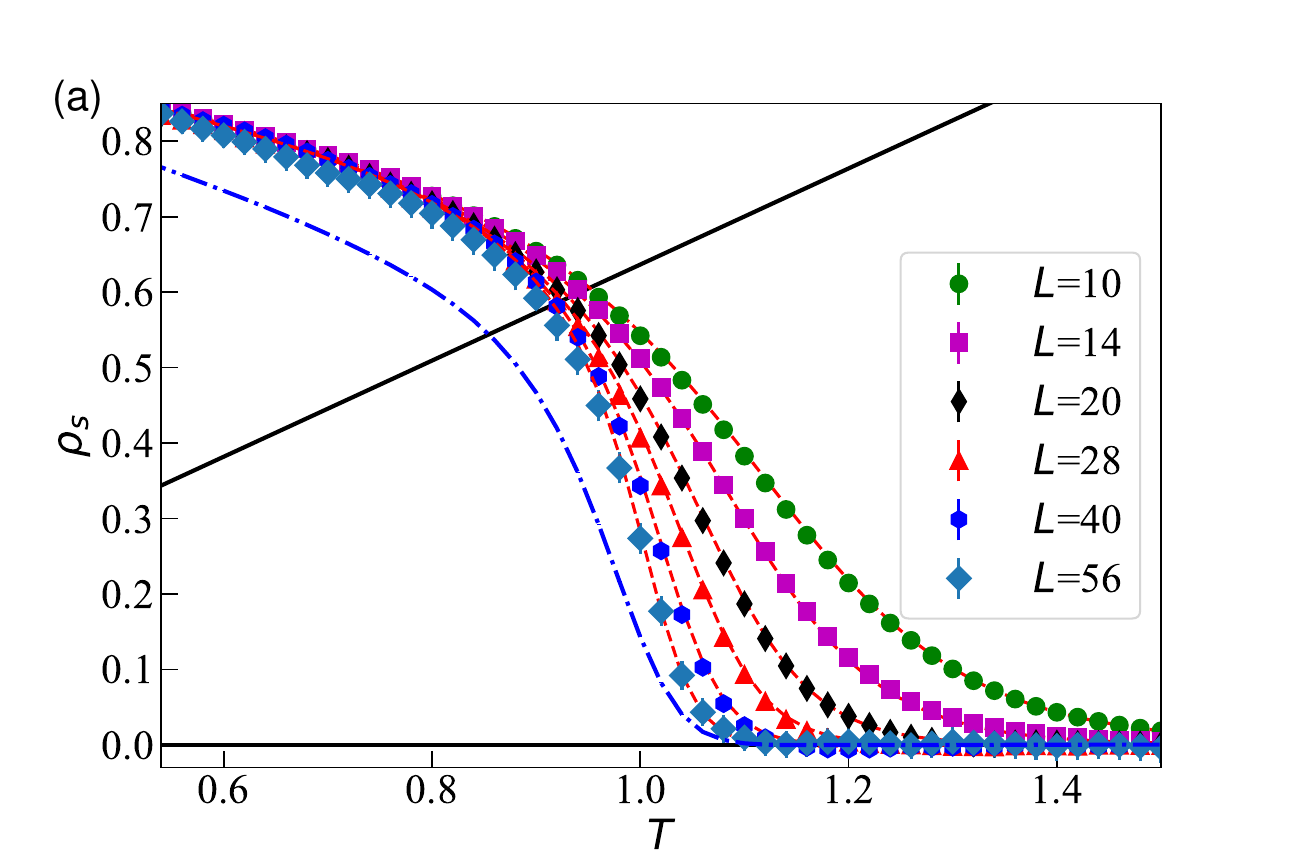}
\includegraphics[width=\columnwidth]{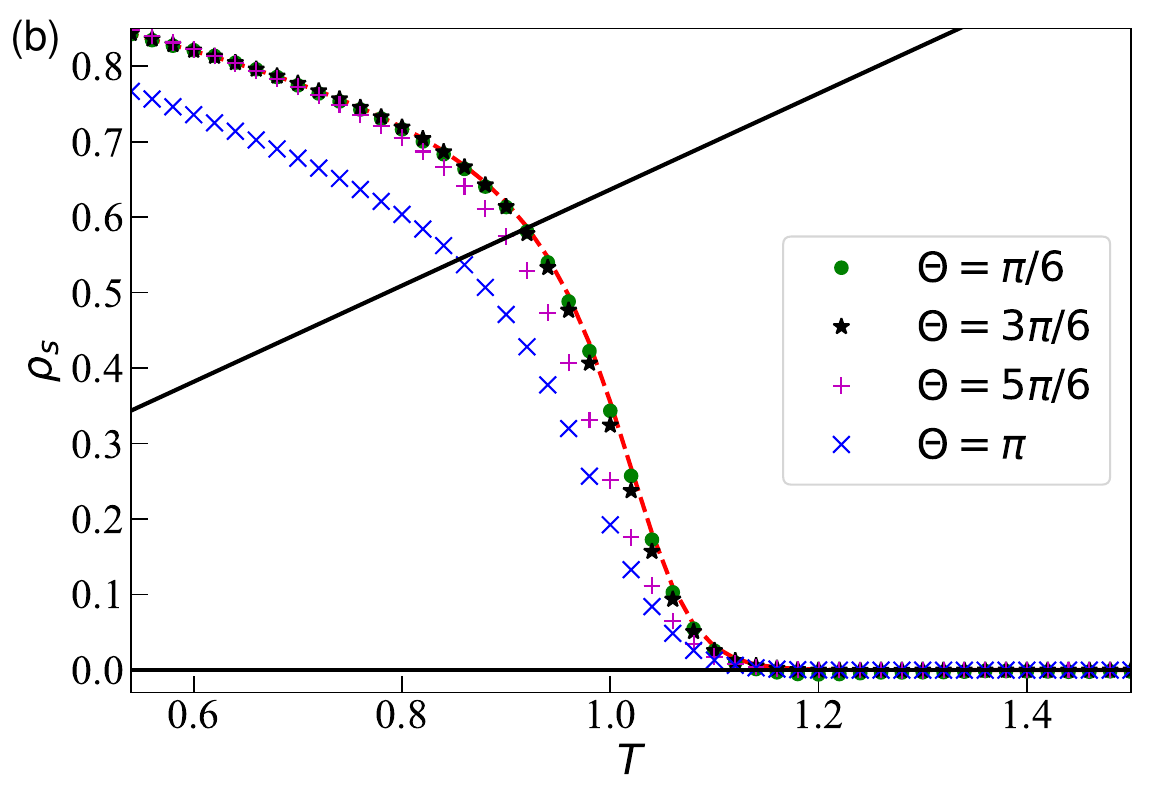}
\caption{(a) Helicity modulus $\rho_s(\Theta)$ as a function of temperature $T$ for different system sizes $L$.
The filled symbols show $\rho_s$ for a twist of $\Theta=\pi/6$, obtained via Eqs.\ (\ref{eq_stiff}) and (\ref{eq_freeint}). The dotted lines show $\rho_{s0}$ for an infinitesimal twist measured in the untwisted system via Eq.\ (\ref{eq:infinitesimal_stiffness}). The straight solid line corresponds to $\rho_s = 2T/\pi$, it intersects the stiffness curves at temperatures $T^*(L)$. The critical temperature $T_c$ is obtained by extrapolating $T^*(L)$ to the thermodynamic limit according to Eq.\ (\ref{eq:Tcextapolate}). This extrapolation gives $T_c=0.891(7)$. In contrast, the helicity modulus values for a twist of $\Theta=\pi$, shown as dash-dotted line for $L=56$, are significantly lower than the other data.
(b)  Helicity modulus $\rho_s(\Theta)$ as a function of temperature $T$ at fixed $L=40$ for different twist angles $\Theta$. The statistical errors in both panels are comparable to or smaller than the symbol sizes.
}
\label{fig_stiffcompare}
\end{figure}

We now turn to our results for the response of the two-dimensional XY model to various twists in the boundary conditions.
Figure \ref{fig_stiffcompare}(a) shows the helicity modulus $\rho_s(\Theta)$ as a function of temperature for
an infinitesimal twist as well as a twist of $\Theta=\pi/6$.
For the infinitesimal twist, $\rho_s$ is measured in the untwisted system via eq.\ (\ref{eq:infinitesimal_stiffness}). For a finite twist, $\rho_s$ is obtained from the free energy difference between simulations with twisted and untwisted boundary conditions, as explained in Sec.\ \ref{sec:MC}. The resulting $\rho_s$ values for the infinitesimal twist and $\Theta=\pi/6$ agree within their statistical errors, giving us additional confidence in our numerical approach.

The helicity modulus curves can be used to find the critical temperature. In the thermodynamic limit, $\rho_s$ vanishes in the disordered phase whereas it is nonzero in the quasi long-range ordered phase. The BKT transition is marked by a universal jump in $\rho_s$.
Using the Kosterlitz-Nelson relation, $T_c$ can be identified by the intersection of the infinite-system $\rho_s$ vs.\ $T$ curve with a straight line of slope $2/\pi$ \cite{nelson1977universal}.
As the correlation length increases exponentially for $T \rightarrow T_c$ at a BKT transition, finite-size corrections take a logarithmic form. Thus, $T_c$ is found by extrapolating $T^*(L)$, the temperature at which the $\rho_s$ vs.\ $T$ curve for size $L$ intersects the line of slope $2/\pi$, according to
\begin{equation}
\label{eq:Tcextapolate}
T^*(L) = T_c + \frac{A}{\ln^2(bL)}
\end{equation}
where $A, b$ are non-universal fitting parameters. We find $T_c=0.891(7)$ from twisted boundary conditions with $\Theta=\pi/6$,
which agrees with high-accuracy results in the literature \cite{ueda2021,hasenbusch1997,hsieh2013finite,Jha2020}.

Figure \ref{fig_stiffcompare}(a) also shows data for $\Theta=\pi$. Unexpectedly, the helicity modulus values resulting from eq.\ (\ref{eq_stiff}) in this case are significantly below the values for smaller $\Theta$. This is further illustrated in Fig.\ \ref{fig_stiffcompare}(b) which compares the helicity modulus for different twist angles $\Theta$. Whereas the data for $\Theta=0, \pi/6$, and $\pi/2$ all agree within their error bars, some deviations appear for $\Theta=5\pi/6$ close to the BKT transition. They become more pronounced for $\Theta=\pi$ and persist in the entire quasi-long-range ordered phase.  It is worth emphasizing that this happens even though the layer-to-layer twist $\Theta/L$ remains small compared to unity, justifying the expansion that leads to eq. (\ref{eq_freediff}).
What is the reason for this surprising discrepancy? Arguments put forward in Ref.\ \cite{BrownHelicity} suggest that the chirality of the twist plays an important role.

So far we have considered thermodynamic states that fulfill the global twist $\Theta$ in, say, the clockwise direction by introducing an average clockwise twist of $\Theta/L$ between neighboring layers. However, the same global twist can be achieved by a counter-clockwise twist with local angle of $(2\pi-\Theta)/L$ between consecutive layers. At low temperatures, the additional free energy for a twist in the ``wrong'' direction is much larger than $k_BT$. Therefore, this state is exponentially suppressed. Local twist angles corresponding to higher winding numbers can also be ruled out using the same argument.
However, with increasing temperature, states of both chiralities (and higher winding numbers) will be mixed, leading to an extra entropic contribution to the free energy. Importantly, $\Theta =\pi$ is a special case that allows the mixing of opposite chiralities even at $T=0$.
Let us now quantitatively study the mixing of the states as twist, temperature and system size is varied.

\begin{figure}[t]
\centering
\includegraphics[width=\columnwidth]{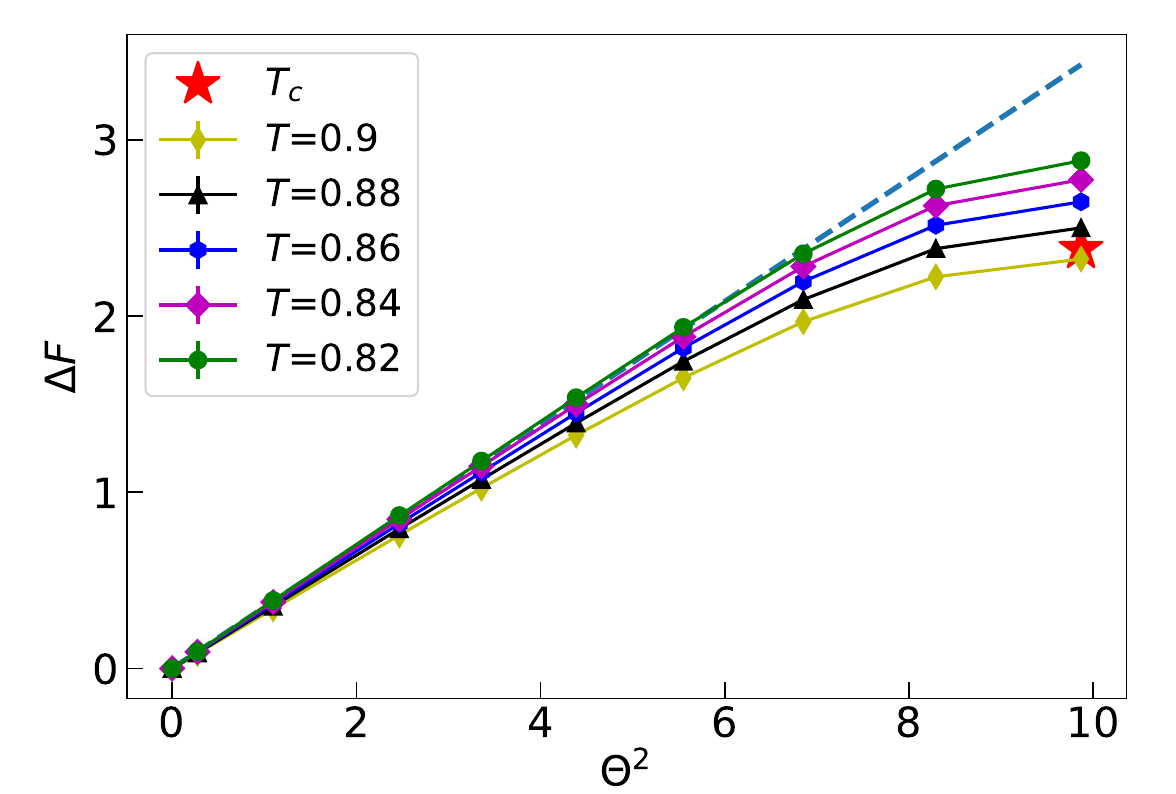}
\caption{Free energy difference $\Delta F$  between systems with periodic and twisted-periodic boundary conditions vs.\ squared twist angle $\Theta^2$ for size $L=40$ and several temperatures close to $T_c$. The dashed line shows a fit of $\Delta F$ for $T=0.82$ to a quadratic $\Theta$ dependence in the range $0 \le \Theta^2 \le (2\pi/3)^2$. $\Delta F$ at $T_c \approx 0.893$, obtained using eq.\ (\ref{eq:ratio_partition}) is marked by the star symbol. Statistical errors are comparable to the symbol size or smaller. }
\label{Fig:dFvsthta2}
\end{figure}

According to eq.\ (\ref{eq_freediff}),  we expect the free energy difference between the twisted and untwisted systems to behave as $\Delta F \propto \Theta^2$, at least as long as higher-order terms in $\Theta/L$ can be neglected. We have studied the $\Theta$ dependence of $\Delta F$ systematically at various temperatures in the quasi long-range ordered phase close to $T_c$; the results are presented in Fig. \ref{Fig:dFvsthta2}.
The figure shows that $\Delta F$ follows the quadratic $\Theta$ dependence up to about $\Theta=2\pi/3$ within the statistical errors of the data, but significant deviations are observed for larger twist angles.

The special case of anti-periodic boundary conditions ($\Theta=\pi$) right at the BKT transition temperature was already studied by Hasenbusch \cite{hasenbusch2009,hasenbusch2005}. He derived an expression for the ratio of the partition functions with periodic and antiperiodic boundary conditions in the two-dimensional XY model. It includes the leading finite-size corrections and reads
\begin{equation}
\label{eq:ratio_partition}
\frac{Z_{\Theta=\pi}}{Z_{\Theta=0}} = 0.08643(1) - \frac{0.1358(1)}{\ln (L) +C}~,
\end{equation}
where the constant $C$ approximately captures contributions from higher order terms. For the purpose of comparing with our Monte Carlo results, we set $C=4.3$ as in Ref.\ \cite{hasenbusch2009}. The free energy difference $\Delta F$ resulting from this formula agrees well with our data, see Fig. \ref{Fig:dFvsthta2}.

We attribute the deviation of $\Delta F$ from the quadratic $\Theta$ dependence to the mixing of states of opposite chirality
which becomes more pronounced as $\Theta$ approaches $\pi$.
The extent of the mixing can be estimated by studying a simple two-state model consisting of a state with clockwise (CW) chirality of the twist and a state with counter-clockwise (CCW) chirality.
At low temperatures and large $L$, the energy associated with the CW state reads $U_{\mathrm{CW}}=-JL^2[1+\cos(\Theta/L)] \approx -2JL^2+J\Theta^2/2$, whereas the energy of the CCW state is given by $U_{\mathrm{CCW}}=-JL^2[1+\cos((2\pi-\Theta)/L)] \approx -2JL^2+J(2\pi-\Theta)^2/2$. At higher temperatures, fluctuations about the perfect spin-wave states become important. Their effect can be approximately captured by replacing $J$ in the above formulas for $U_{\mathrm{CW}}$ and $U_{\mathrm{CCW}}$ by the (renormalized) helicity modulus $\rho_{s}(T)$.
The corresponding canonical probabilities of CW and CCW states read
\begin{equation}
p_{\mathrm{CW}} =\frac 1 Z e^{-\frac{\rho_s \Theta^2}{2T}}~, \quad p_{\mathrm{CCW}} = \frac 1 Z  e^{-\frac{\rho_s (2\pi-\Theta)^2}{2T}}
\label{eq:two_state_probs}
\end{equation}
with
\begin{equation}
Z = e^{-\frac{\rho_s \Theta^2}{2T}} ~+~  e^{-\frac{\rho_s (2\pi-\Theta)^2}{2T}}~.
\end{equation}
As discussed earlier, this implies that the CCW state is exponentially suppressed as $T\to 0$ for any $\Theta<\pi$ whereas both states contribute equally at all temperatures below $T_c$ for $\Theta=\pi$.

\begin{figure}
\centering
\includegraphics[width=\columnwidth]{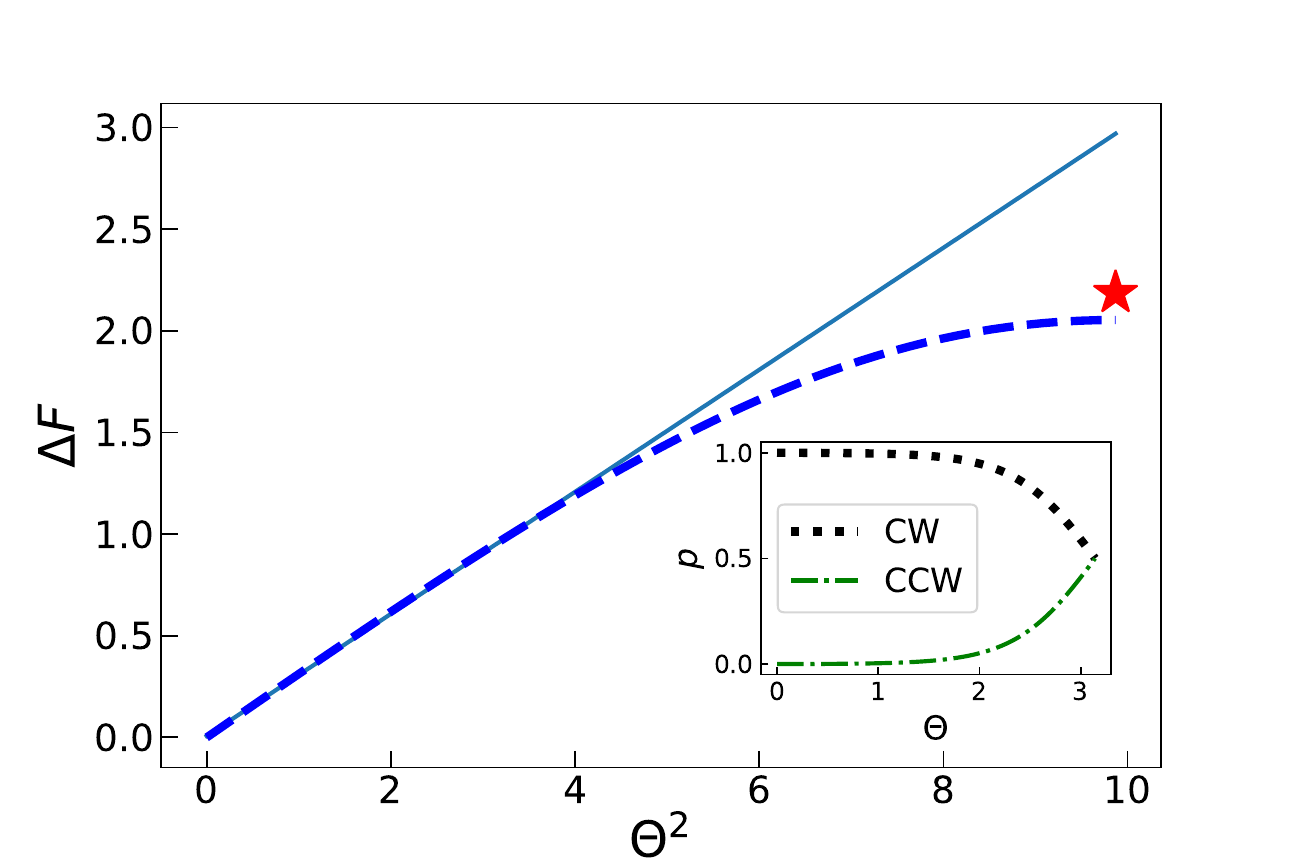}
\caption{Free energy difference $\Delta F$ at the BKT transition temperature $T_c$ between systems with periodic and twisted-periodic boundary conditions, as evaluated using the two-state model (dashed line). The solid line corresponds to a quadratic fit in the range $0 \leq \Theta^2 \leq (2\pi/3)^2$.  The value of $\Delta F$ from eq.\ (\ref{eq:ratio_partition}) in the thermodynamic limit is marked by the star symbol. The canonical probabilities of the two states are shown in the inset as functions of $\Theta$.}
\label{Fig:dFvsthta_toy}
\end{figure}

The helicity modulus right at the BKT transition temperature $T_c$ is known from the Kosterlitz-Nelson relation, $\rho_s(T_c) = \frac{2T_c}{\pi}$.
Inserting this into eq.\ (\ref{eq:two_state_probs}) gives the canonical probabilities shown in the inset of Fig.\ \ref{Fig:dFvsthta_toy}.
The contribution of the CCW state to the mixture increases with $\Theta$ and becomes significant for $\Theta \gtrapprox 2\pi/3$.
The free energy difference $\Delta F$ between systems with periodic and twisted-periodic boundary conditions can also be computed within the two-state model,
\begin{equation}
\Delta F = \langle \Delta U \rangle - TS
\end{equation}
where $S$ is the Von-Neumann entropy. The main panel of Fig.\ \ref{Fig:dFvsthta_toy} shows $\Delta F$ as a function of $\Theta$ at the BKT transition temperature. The deviation of $\Delta F$ from the quadratic dependence on $\Theta$ resembles the corresponding Monte Carlo result in Fig.\ \ref{Fig:dFvsthta2}. Moreover, $\Delta F$ from Eq.\ (\ref{eq:ratio_partition}) in the thermodynamic limit $L=\infty$ (marked by the star symbol) agrees well with the model results. The simple two-state system thus captures the important features present in Fig. \ref{Fig:dFvsthta2}. This supports the notion that the mixing of states with opposite chiralities leads to the deviation of $\Delta F$ from the quadratic $\Theta$ dependence.

We now move beyond the two-state model and quantify the mixing of chiralities in the Monte Carlo data. To determine the excess free energy due to the mixing, we first fit a quadratic function to $\Delta F(\Theta)$ in the range $0 \le \Theta \le 2\pi/3$
(separately for each system size).
Denoting the fit function by $\Delta F_{\mathrm{fit}}(L,T,\Theta)$, we define the excess free energy $\delta \Delta F$
at twist angle $\pi$ as the difference between $\Delta F_{\mathrm{fit}}(L,T,\pi)$ and the $\Delta F_{\mathrm{MC}}(L,T,\pi)$ obtained from MC,
\begin{equation}
\label{eq:deltadeltaF}
\delta \Delta F(L,T,\pi) = \Delta F_{\mathrm{fit}}(L,T,\pi) - \Delta F_{\mathrm{MC}}(L,T,\pi).
\end{equation}
In other words, $\Delta F_{\mathrm{fit}}$ is the free energy cost of the twist expected if only states of one chirality contribute, whereas $\delta \Delta F$ captures the additional free energy due to chirality mixing.

\begin{figure}[t]
\centering
\includegraphics[width=\columnwidth]{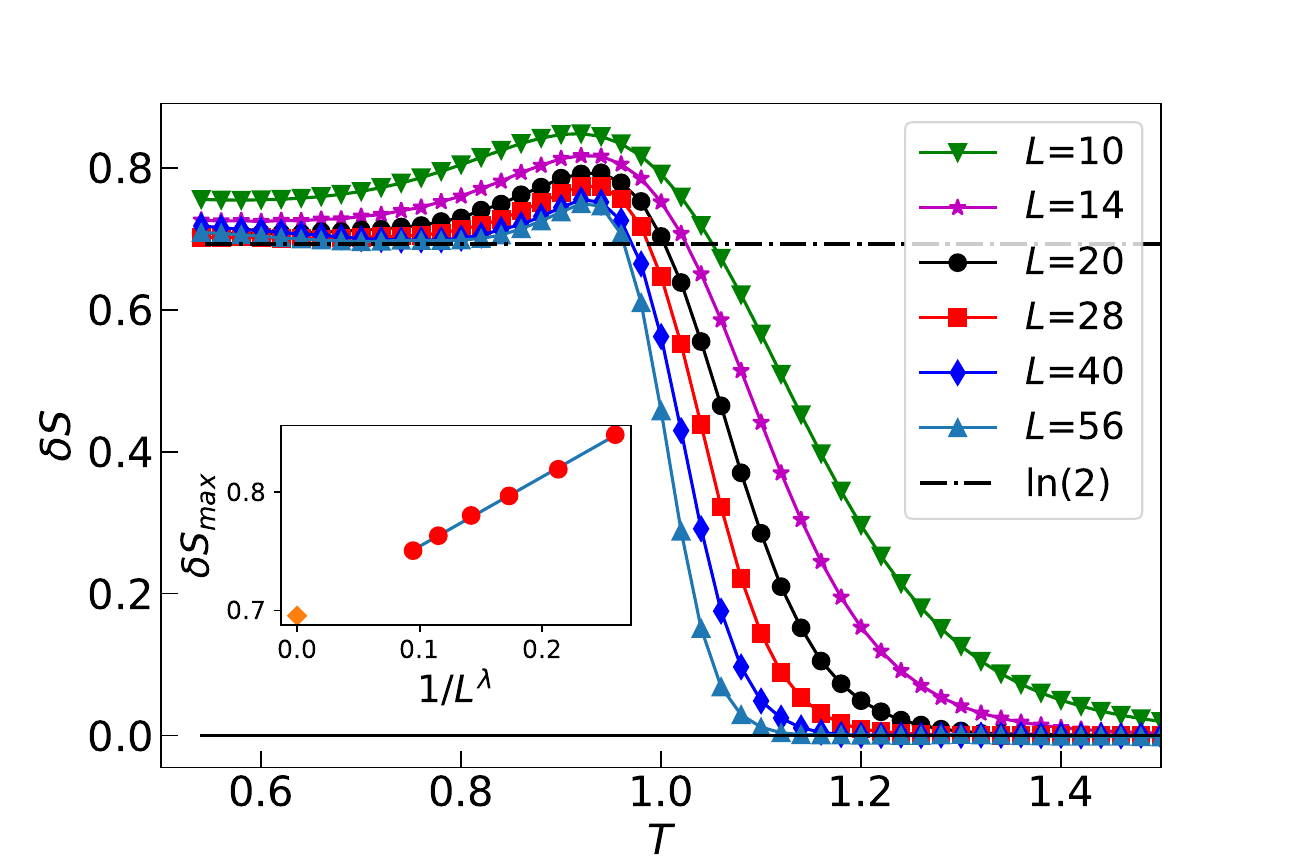}
\caption{Excess entropy $\delta S$ due to the mixing of chiralities vs.\ temperature $T$ for different system sizes. The solid horizontal line corresponds to $\delta S =\ln(2)$.
The inset shows the extrapolation of the maximum value of each $\delta S$ vs.\ $T$ curve to infinite system size. The extrapolation gives $\delta S_{\textrm{max}}(L=\infty)=0.69(1)$ and agrees with $\ln(2)$.}
\label{ddSvsT}
\end{figure}

As the excess free energy is expected to be entropic in nature, we define $\delta S = \delta \Delta F /T$ as the excess entropy due to the mixing of chiralities. Our numerical results for the excess entropy
are presented in Fig.\ \ref{ddSvsT}.
The figure shows that  $\delta S$ approaches zero with increasing system size for high temperatures while it appears to approach a constant for low temperatures. In fact, the figure suggests that $\delta S$ approaches a step function in the thermodynamic limit. To determine the step height and position, we extrapolate the peak value $\delta S_{\textrm{max}}$  of the $\delta S$ vs.\ $T$ curves using,
\begin{equation}
\delta S_{\textrm{max}}(L) = \delta S_{\textrm{max}}(\infty) + a L^{-\lambda}
\end{equation}
where $a, \lambda$ are fitting parameters. (We determine $\delta S_{\textrm{max}}$ as an extremum of a quadratic curve fitted in the vicinity of the peak.)
The extrapolation is shown in the inset of Fig.\ \ref{ddSvsT} and yields $\delta S_\infty = 0.69(1) \approx \ln(2)$, in agreement with the expectation of contributions from two degenerate states.
Additionally, an extrapolation of the temperature at which the  $\delta S$ vs.\ $T$ curves cross the $\delta S = \ln (2)$ line
matches with $T_c$ within the error bars. Thus, our numerical data extrapolate to $\delta S = \ln(2) \Theta(T_c-T)$ where
$\Theta$ is the Heavyside step function. We note that the arguments predicting the excess $\ln(2)$ entropy due to the chirality mixing do not include states with higher winding numbers. These states are known to renormalize the helicity modulus in two dimensions
\cite{ProkofevSvistunov00,hsieh2013finite}, but the effect is tiny and only visible in high-accuracy simulations beyond our numerical precision.

\begin{figure}[t]
\centering
\includegraphics[width=\columnwidth]{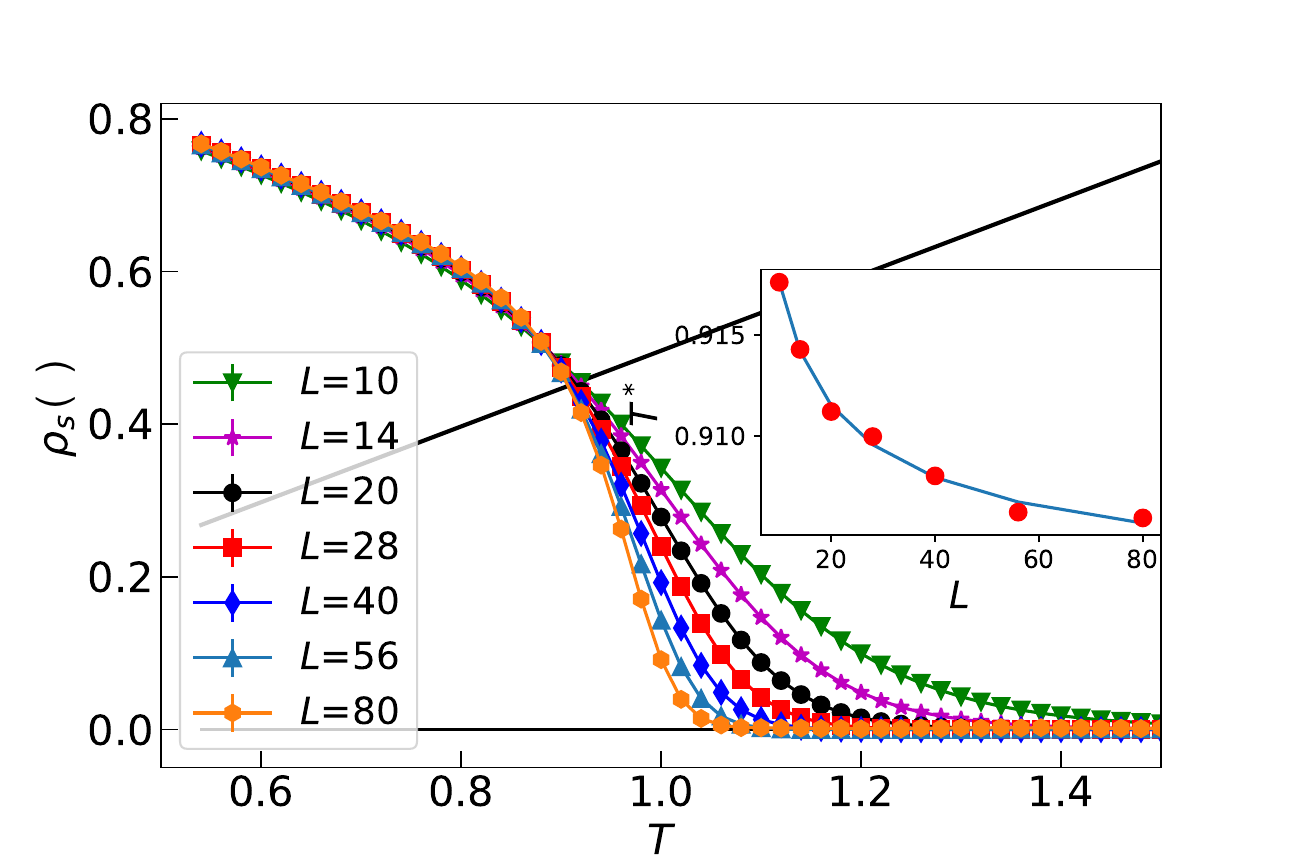}
\caption{Helicity modulus $\rho_s(\pi)$ for a twist of $\Theta=\pi$ (anti-periodic boundary conditions). The inset shows the intersection temperatures $T^*$ of the helicity modulus curves with the modified Kosterlitz-Nelson line of slope $2/\pi - 2~\ln(2)/\pi^2$. An extrapolation according to eq.\ (\ref{eq:Tcextapolate}) yields $T_c =0.897(2)$.} \label{fitstiff}
\end{figure}

The excess entropy due to the chirality mixing reduces the free energy cost of a $\pi$ twist by $T \ln(2)$. Consequently,
the helicity modulus $\rho_s(\pi)$ computed from eq.\ (\ref{eq_stiff}) is reduced by $2 \ln(2) T /\pi^2$ compared to the infinitesimal twist value
arising from eq.\ (\ref{eq_stiff_inf}).
This explains the observation in Fig.\ \ref{fig_stiffcompare} of a lower helicity modulus for the $\pi$ twist. Consequently,
if one wishes find the critical temperature from simulations employing a $\pi$ twist (i.e., anti-periodic boundary conditions),
the reduction of $\rho_s$ has to be accounted for. This can be achieved by adding the correction to the Kosterlitz-Nelson relation by
changing the slope of the line crossing the $\rho_s$ curves from $2/\pi$ to $2/\pi - 2~\ln(2)/\pi^2$.
The resulting analysis is presented in Fig.\ \ref{fitstiff} which shows the helicity modulus for a $\pi$ twist as a function of temperature for different system sizes.
To find $T_c$ we extrapolate, using eq.\ (\ref{eq:Tcextapolate}), the intersection temperature of the $\rho_s$ curves with a line of modified slope, and obtain $T_c =0.897(2)$. This is in reasonable agreement with the literature value \cite{hsieh2013finite}.

\subsection{Breaking of the chiral symmetry}

In Sec.\ \ref{sec:results-free-energy}, we have established that, for an imposed twist $\Theta=\pi$, the thermodynamic ensemble contains an equal mixture of states with clockwise and counter-clockwise local twists.
Now we explore the question whether a single given macroscopic system fluctuates between clockwise and counter-clockwise orientations or whether it spontaneously breaks the chiral symmetry below $T_c$ by
freezing into one orientation. As the classical XY Hamiltonian does not contain any dynamic terms, the answer to this question will depend on the assumed dynamics of the system. Here, we focus on dissipative local model-A
dynamics according to the Hohenberg-Halperin classification \cite{HohenbergHalperin77}, i.e., a purely relaxational dynamics without conservation laws. However, the results should hold qualitatively for other local dynamics as well.

To analyze chiral fluctuations, we define the bulk chirality $h$ via the vector product between nearest-neighbor spins along the direction of the twist. Assuming a twist in the $x$-direction, $h$ is given by
\begin{equation}
\label{eq:chirality}
h =  \frac{1}{L}  \sum_{\langle ij \rangle_x} \mathbf{\hat k} \cdot \left({\mathbf{S}}_{i} \times {\mathbf{S}}_{j} \right)
\end{equation}
where site $j$ is the nearest neighbor of site $i$ in the positive $x$ direction, and $\mathbf{\hat k}$ is the unit vector in the $z$-direction.
This means positive $h$ correspond to a counter-clockwise twist and negative $h$ correspond to a clockwise twist.

\begin{figure}
\centering
\includegraphics[width=\columnwidth]{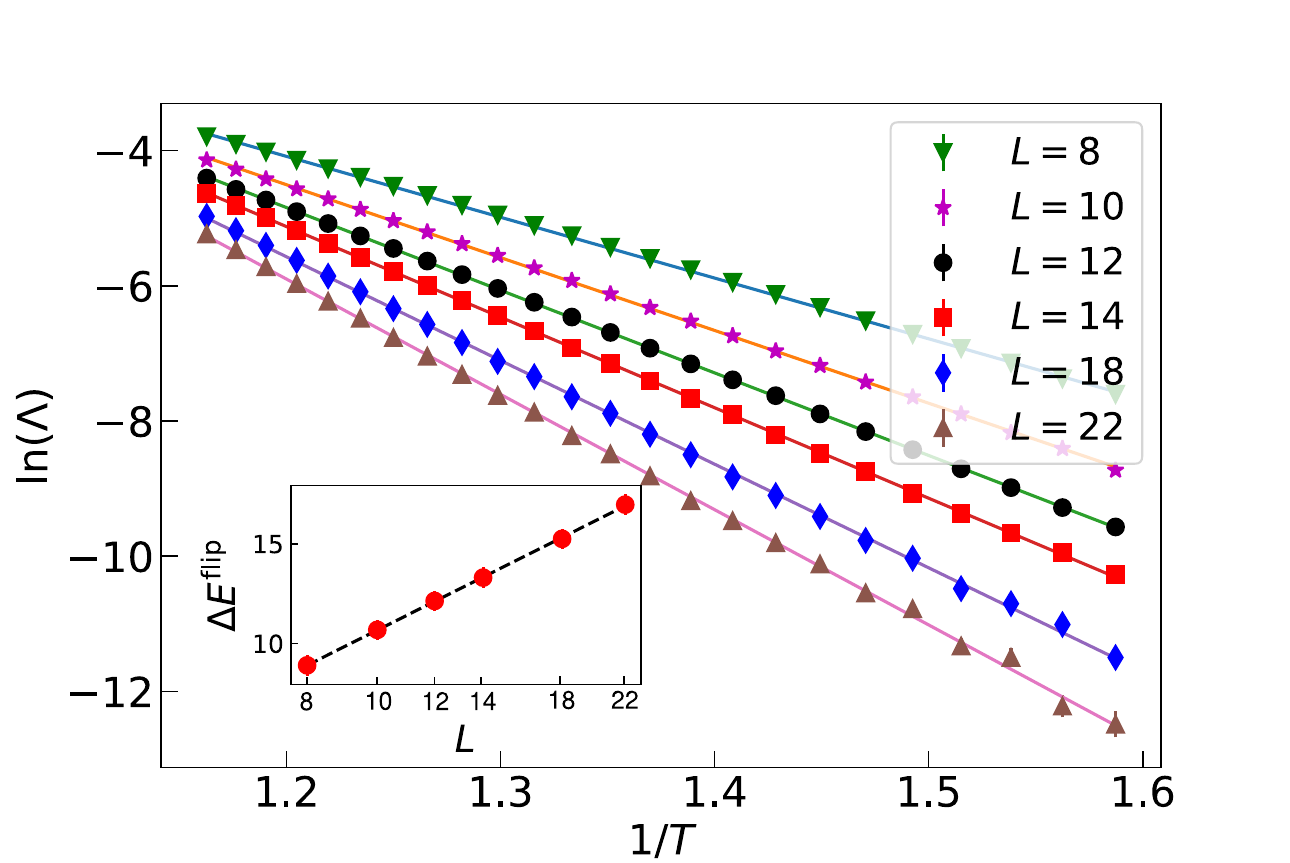}
\caption{Flip rate $\Lambda$ of the bulk chirality $h$ as a function of the inverse temperature $1/T$ for different system sizes $L$. The solid lines are fits to the ansatz  (\ref{eq:fliprate}). The resulting system size dependence of the activation energy $\Delta E^{\mathrm{flip}}$ is shown in the inset in a semi-logarithmic plot. The dotted line is a fit to the function, $\Delta E^{\mathrm{flip}} = E_0 \ln(k_0 L)$, giving fit parameters $E_0=7.91(6)$ and $k_0=0.39(1)$. }
\label{fliprate}
\end{figure}

We monitor the time evolution of the bulk chirality $h$ in long Monte Carlo runs that perform only local Metropolis updates (implementing model-A dynamics). Specifically, we measure the rate  $\Lambda$ at which the chirality changes sign during such a simulation.
Figure \ref{fliprate} shows the flip rate $\Lambda$ as a function of temperature for different system sizes observed for runs of $1.2 \times 10^6$ Metropolis sweeps averaged over $400$ samples.
The data clearly suggest an exponential dependence of the flip rate on the inverse temperature.
Indeed, if chirality flips are governed by activation over an energy barrier $\Delta E^{\mathrm{flip}}$, the flip rate is expected to follow the ansatz
\begin{equation}
\label{eq:fliprate}
\Lambda = c T^b \exp(-\Delta E^{\mathrm{flip}}/T)~.
\end{equation}
where $b$ and $c$ are fit parameters. Figure \ref{fliprate} demonstrates that the observed flip rates indeed follow this ansatz within their statistical errors.
The activation energies $\Delta E^{\mathrm{flip}}$ obtained from the fits in Fig.\ \ref{fliprate} are presented in the inset as a function of the system size
in a semi-logarithmic plot. The data are well described by the logarithmic function
\begin{equation}
\Delta E^{\mathrm{flip}} = E_0 \ln(k_0L)
\label{eq:flipenergy}
\end{equation}
with $E_0=7.91(6)$ and $k_0=0.39(1)$.  This logarithmic dependence suggests that flips of the bulk chirality are facilitated by a vortex mechanism.

\begin{figure}
\centering
\includegraphics[width=\columnwidth]{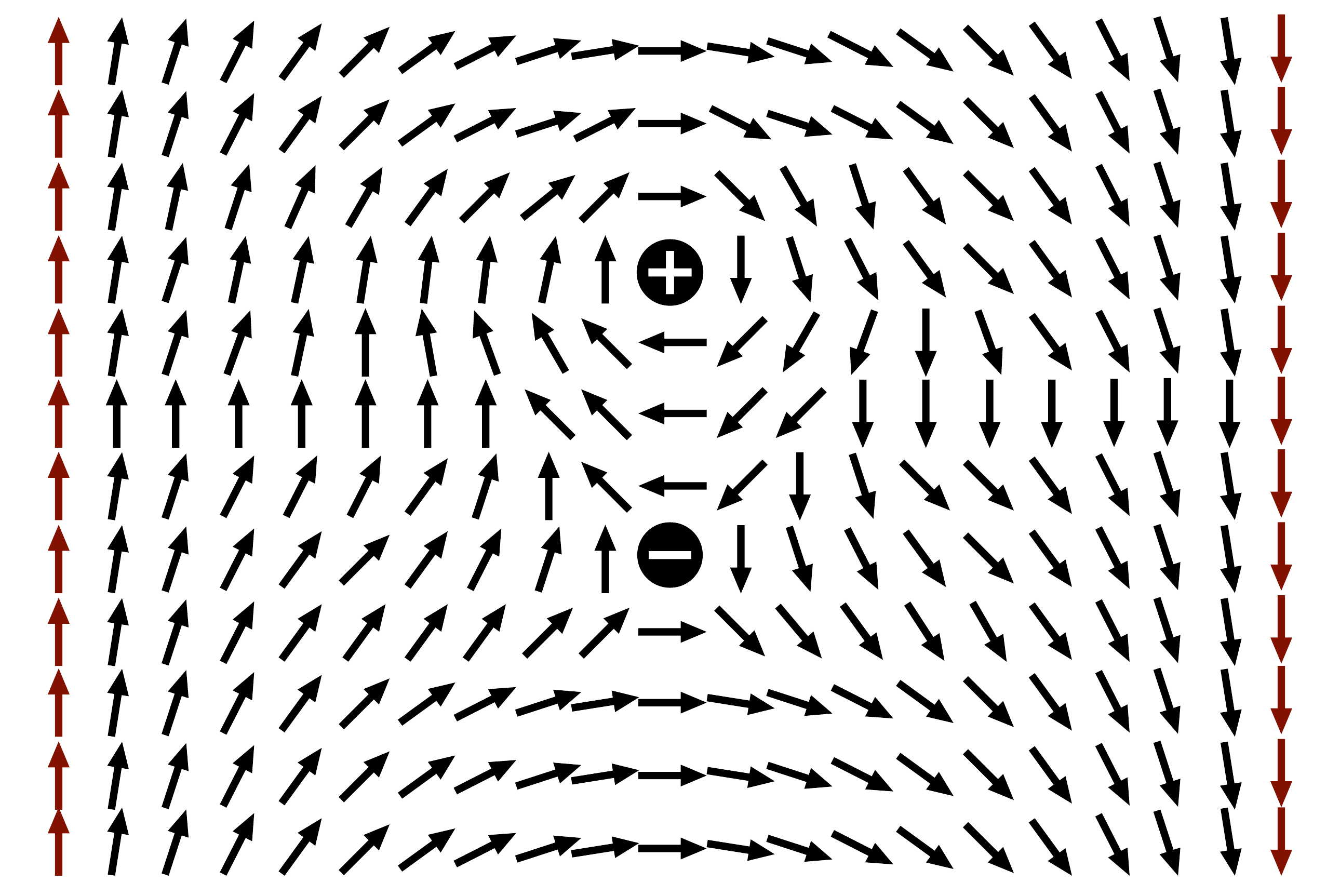}
\caption{Schematic illustrating how a vortex-antivortex pair can flip the bulk chirality. For details see text.
}
\label{vortices_sketch}
\end{figure}

Figure \ref{vortices_sketch} illustrates how the creation and annihilation of a vortex-antivortex pair can reverse the sign of the bulk chirality.
Imagine a system with fixed boundaries in the $x$-direction that impose a $\pi$ twist and periodic boundary conditions in the $y$-direction.
Initially, the system features a uniform clockwise twist to satisfy the boundary conditions. When a vortex-antivortex pair is introduced as shown in the figure,
the spins between the vortices now rotate in the opposite direction. Now imagine that the + vortex travels upwards across the (periodic) boundary, approaches
the $-$ vortex from below, and finally annihilates. The end result is a state with a uniform counter-clockwise twist, i.e., the chirality $h$ has changed sign.
The activation energy of this process is given by the energy of a vortex-antivortex pair at the largest distance $L/2$. In the continuum limit,
the energy of a vortex-antivortex pair at distance $D$ is given by $E_{\mathrm{pair}}= E_{\mathrm{core}} + 2 \pi J \ln\left(D/a)\right)$ where
$E_{\mathrm{core}}$ is the vortex core energy and $a$ the core size. Neglecting the core energy (which can be formally absorbed by a shift of $a$),
a comparison with eq.\ (\ref{eq:flipenergy}) gives a reasonable agreement. This
indicates that the leading contribution to $\Delta E^{\mathrm{flip}}$ comes from the formation of a vortex-antivortex pair. Deviations
can be attributed to finite-size effects, contributions from other flip processes and uncertainties in the sequence of fits leading to
eq.\ (\ref{eq:flipenergy}) \footnote{If one considers open rather than periodic boundary conditions in the direction perpendicular to the twist,
a single vortex moving across the system is sufficient to flip the bulk chirality. Test simulations with open boundary conditions lead to the same
logarithmic dependence (\ref{eq:flipenergy}) of the activation energy $\Delta E^{\mathrm{flip}}$ but with prefactor  $E_0 = 3.7(1)$, indicating the formation
of a single vortex.}.

The results of this section demonstrate that the energy required to flip the bulk chirality diverges with system size in the thermodynamic limit. The divergence is only
logarithmic in $L$ in contrast to the case of a domain wall whose energy would diverge linearly in $L$ in two dimensions. Nonetheless, it implies that a
macroscopic system will not fluctuate between states of opposite bulk chirality during its time evolution (for local dynamics). In other words, the chiral
symmetry is spontaneously broken below the BKT transition.

\section{Conclusion}
\label{sec:conclusions}

To summarize, in this paper we have studied the effects of finite (non-infinitesimal) twists in the boundary conditions on a two-dimensional classical ferromagnetic XY model.
The system's response to the twist has been studied by a direct evaluation of the free energy by means of large-scale Monte Carlo simulations.
We have found that in the quasi long-range ordered phase below the BKT transition, the free energy cost of a non-infinitesimal twist deviates from the expected quadratic dependence on the twist angle.
In the case of a $\pi$ twist (anti-periodic boundary conditions), the mixing of states of opposite chiralities causes an excess entropy contribution of $\ln(2)$ that lowers
the free energy cost of the twist. Thus, if the helicity modulus is calculated from the free energy of a $\pi$-twisted system (using eq.\ (\ref{eq_stiff_inf})), its
value is reduced compared to the usual helicity modulus obtained from an infinitesimal twist. We note that our results have been obtained by comparing periodic and twisted-periodic boundary conditions. However, all our arguments are expected to hold as well for systems in which the boundary spins are held at fixed angles.
We also note that our discussion has been based on the phase
representation of the XY model, but the same results are expected in other representations \cite{villain1975,jose1977renormalization,cardy1980,vallat1994}
provided they correctly reflect the chiral symmetry of the boundary conditions.

A macroscopic $\pi$-twisted system in the quasi long-range ordered phase spontaneously breaks the chiral symmetry between states with clockwise and counter-clockwise local twists
(at least in the case of local dynamics). This implies that experiments on a single macroscopic system cannot observe the $\ln(2)$ entropy contribution from chirality mixing and the
corresponding reduction of the apparent helicity modulus value. The excess entropy is expected to be observable in mesoscopic systems that are small enough to fluctuate between states
with opposite chirality and in situations involving not just a single system but an entire (thermodynamic) ensemble.

Chirality mixing is important for computer simulations for at least two reasons. First, efficient simulations often involve nonlocal algorithms (such as the Wolff cluster
algorithm) that can freely flip the bulk chirality of the system. Moreover, system sizes in computer simulations are sometimes not very large so that even algorithms with local
updates (such as the Metropolis algorithm) may allow the system to fluctuate between different chiralities.
Free energies of $\pi$-twisted systems computed in such simulations contain the $-T\ln(2)$ entropic free energy contribution due to chirality mixing.
To compute the usual (infinitesimal-twist) helicity modulus $\rho_{s0}$ from simulations with a $\pi$ twist, this entropic contribution needs to be removed.
When determining the BKT transition temperature from the helicity modulus data, one can, alternatively, modify the Kosterlitz-Nelson relation as discussed in Sec.\
\ref{sec:results-free-energy}.

The effects of twisted boundary conditions on an XY model in thin-film geometry were also studied in Ref.\ \cite{bergknoff2011} using a mean-field theory. For a twist of $\Theta=\pi$, this work finds an additional singularity of the free energy below the bulk critical temperature. It is induced by the boundary conditions and leads to kink in the Casimir force. In our calculations, we do not observe such a singularity. We believe that this may stem from the fact that the mechanism proposed in Ref.\ \cite{bergknoff2011}, viz., a competition between a rotational state and a planar state to fulfill the boundary conditions, requires soft spin variables and does not hold for the hard spins (of fixed unit length) of our XY model.
Alternatively, the lack of additional singularity might be because states with both directions of rotation are simultaneously realized in our simulations at all twist angles, whereas in Ref.\ \cite{bergknoff2011}, only one direction is considered in the calculations. 

The partition function of a quantum system in $d$ dimensions can often be mapped onto that of a classical system in $d+1$ dimensions. Our results thus apply to a one-dimensional (particle-hole symmetric) quantum rotor model with twisted boundary conditions. Specifically, our findings imply that in a finite-size rotor model the states with opposite chirality hybridize. However in the thermodynamic limit, the ground state is doubly degenerate because the hybridization goes to zero and the tunneling time between states of opposite chirality diverges with system size. Analogous results are expected in a field-theoretic approach based on the sine-Gordon model \cite{Giamarchi04,Lukyanov98}.

The present paper has focused on two dimensions. Two dimensions are a special case because the free energy cost of a twist is independent of system size, see eq.\ (\ref{eq_freediff}). Thus, the excess entropy discussed above makes a non-negligible contribution even in the thermodynamic limit. In a higher dimensional XY model, the excess entropy would still take the value ln(2) for a $\pi$ twist. However, its contribution to the helicity modulus, $2\ln(2) L^{2-d}/\pi^2$, would vanish in the thermodynamic limit.

Our work also relates to some of the questions raised by Brown and Ciftan \cite{BrownHelicity}. They studied the effects of twisted-periodic and anti-periodic boundary conditions on a three-dimensional classical Heisenberg model and discussed the notion of mixing states of different chiralities in response to twisted-periodic boundary conditions. However, they analyzed the internal energy cost of the twist rather than the free energy cost. The authors report deviations from a quadratic twist angle dependence of the internal energy cost somewhat similar to what we find for $\Delta F$, but the magnitude of the deviation in their case is much larger than $T \ln(2)$. Moreover, the internal energy cost (as opposed to $\Delta F$) is not expected to  contain the entropy due to the mixing of chiralities. This suggests that the deviations from a quadratic twist angle
dependence of the internal energy cost in Ref.\ \cite{BrownHelicity} have a different origin. Specifically, Heisenberg spins can reduce the energy cost of twisted-periodic boundary conditions by tilting out of the plane in which the twist is applied. In contrast, they cannot avoid antiperiodic boundary conditions, in agreement with the fact that the energy cost of anti-periodic boundary conditions in Ref.\ \cite{BrownHelicity} is much larger than that of even the largest twist angles. A quantitative analysis
of the effects of twisted-periodic boundary conditions in the Heisenberg case remains a task for the future.

Our results have potential implications for experiments proposed to detect the BKT phase transitions \cite{troncoso2020}, in which anti-parallel external fields would be used to study charge-current cross correlations. Moreover, in discrete spin systems such as the $q-$state clock model \cite{Kumano2013}, twists are necessarily non-infinitesimal, and corrections to the infinitesimal-twist helicity modulus must be considered.

\begin{acknowledgments}
We acknowledge support from the National Science Foundation under grant nos. DMR-1506152, DMR-1828489, OAC-1919789.
Simulations were performed on the Foundry and Pegasus Clusters at Missouri University of Science and Technology.
We would also like to thank John Chalker, Eduardo Fradkin, and Martin Hasenbusch for helpful correspondence and W. Joe Meese for valuable discussions.
\end{acknowledgments}
\appendix
\section{Helicity modulus for an infinitesimal twist}
\label{sec_appndx}
In this Appendix, we derive the expression (\ref{eq:infinitesimal_stiffness}) for the second derivative of the free energy on the twist angle. This expression allows the evaluation of helicity modulus for an infinitesimal twist in terms of correlation functions of the untwisted system. The derivation follows Refs.\ \cite{Caffarel_1994,HrahshehBarghathiVojta11}.

Consider an XY magnet in $d$ dimensions, of linear size $L$.
Applying twisted boundary conditions (BC) along $x$ axis implies that the spins at $x=0$ make an angle $\Theta$ w.r.t. the spins at $x=L$.
The change in free energy due to the twist can be parameterized as
\begin{equation}
\Delta F = \frac{1}{2} \rho_s V \left( \frac{\Theta}{L} \right)^2,
\end{equation}
where $V=L^d$. In the limit of $\Theta \rightarrow 0$,
\begin{equation}
\rho_s = \frac{L^2}{V} \left( \frac{\partial^2F}{\partial\Theta^2} \right)_{\Theta=0}.
\end{equation}
The second derivative of the free energy can be evaluated by treating $\Theta$ as a parameter of the partition function.

To this end, we start from the Hamiltonian
\begin{equation}
H = - \sum_{<ij>} J \cos (\phi_i - \phi_j)
\end{equation}
with boundary conditions  $\phi_i = 0$ at $x_i=0$ and $\phi_i = \Theta$ at $x_i=L$. We now perform a variable transformation
$\psi_i = \phi_i - \Theta x_i / L$. Note that $\psi_i$ has untwisted BC, $\psi_i =\phi_i = 0$ at $x_i=0$ and $x_i=L$.
The dependence on the twist angle $\Theta$ has been moved to the Hamiltonian
\begin{equation}
H = - \sum_{<ij>} J \cos (\psi_i +\Theta \frac{x_i}{L} - \psi_j -\Theta \frac{x_j}{L})~,
\end{equation}
\begin{equation}
H = - \sum_{<ij>} J \cos (\psi_i - \psi_j -\frac{\Theta}{L}(x_i-x_j))~.
\label{eq_H_psi}
\end{equation}
The evaluation of $(\partial^2 F/\partial \Theta^2)$ is now straight forward, starting from
\begin{equation}
F = -k_B T\ln (Z({\Theta})) = -k_B T \ln \left( \textrm{Tr} e^{-\beta H }\right)~.
\end{equation}
First,
\begin{align*}
\frac{\partial F}{\partial \Theta} =& -k_B T \frac{1}{Z} \left( \frac{\partial Z}{\partial \Theta} \right)  \\
\frac{\partial F}{\partial \Theta} =& \frac{1}{Z} \textrm{Tr} \left( \frac{\partial H}{\partial \Theta} e^{-\beta H} \right) =\left\langle \frac{\partial H}{\partial \Theta} \right\rangle.
\end{align*}
Then,
\begin{align*}
\frac{\partial^2 F}{\partial \Theta^2} &= \frac{\partial}{\partial \Theta} \left( \frac{1}{Z} \textrm{Tr} \left( \frac{\partial H}{\partial \Theta} e^{-\beta H} \right)\right) \\
\frac{\partial^2 F}{\partial \Theta^2} &= \beta \left\langle \frac{\partial H}{\partial \Theta}\right\rangle^2 +\left\langle \frac{\partial^2 H}{\partial \Theta^2} \right\rangle -\beta \left\langle \left(\frac{\partial H}{\partial \Theta}\right)^2 \right\rangle.
\end{align*}
Each derivative in the above equation can be evaluated from Eq.\ (\ref{eq_H_psi}).
\begin{align}
\left(\frac{\partial H}{\partial \Theta} \right)_{\Theta=0} =& \frac{1}{L} \sum_{<ij>}J \sin (\psi_i-\psi_j)(x_i-x_j)\\
\left(\frac{\partial^2 H}{\partial \Theta^2} \right)_{\Theta=0} =& \frac{1}{L^2} \sum_{<ij>}J \cos (\psi_i -\psi_j) (x_i -x_j)^2.
\end{align}
Collecting all the terms,
\begin{equation}
\left(\frac{\partial^2 F}{\partial \Theta^2} \right)_{\Theta=0} = \beta \left\langle \frac{\partial H}{\partial \Theta}\right\rangle^2_{\Theta=0} + \left\langle \frac{\partial^2 H}{\partial \Theta^2}\right\rangle_{\Theta=0} - \beta \left\langle \left(\frac{\partial H}{\partial \Theta}\right)^2\right\rangle_{\Theta=0}.
\end{equation}
The first term vanishes due to symmetry, but
\begin{align*}
\left(\frac{\partial^2 F}{\partial \Theta^2} \right)_{\Theta=0} =&\frac{1}{L^2} \left\langle \sum_{<ij>}J  \cos (\psi_i -\psi_j)(x_i -x_j)^2\right\rangle  \\
-& \frac{\beta}{L^2} \left\langle \sum_{<ij>} \left\lbrace J \sin (\psi_i -\psi_j) (x_i -x_j)\right\rbrace^2 \right\rangle
\end{align*}
where the thermodynamic averages $\langle ... \rangle$ are defined for $\Theta=0$.
In vector notation, $\cos (\psi_i-\psi_j) = \vec{S}_i \cdot \vec{S}_j,~ \sin (\psi_i -\psi_j) = -\hat{k}\cdot (\vec{S}_i \times \vec{S}_j)$.
Thus,
\begin{align*}
~~~~\left(\frac{\partial^2 F}{\partial \Theta^2} \right)_{\Theta=0} &= \frac{1}{L^2} \sum_{<ij>} J \langle \vec{S}_i \cdot \vec{S}_j \rangle (x_i-x_j)^2 \\
&-\frac{\beta}{L^2}\left\langle \left\lbrace \sum_{<ij>} J (\hat{k}\cdot (\vec{S}_i \times \vec{S}_j)(x_i-x_j)) \right\rbrace^2 \right\rangle.
\end{align*}
The $\Theta \rightarrow 0$ limit thus allows a simple evaluation of $({\partial^2 F}/{\partial \Theta^2}) $ and  of the helicity modulus $\rho_s$. This completes the derivation of Eq.\ (\ref{eq:infinitesimal_stiffness}).

\bibliography{qstaterefs}

\end{document}